\documentclass[sigconf]{acmart}
\usepackage{booktabs} 
\usepackage{graphicx}

\usepackage{color,xcolor}
\usepackage{lipsum}                     
\usepackage{xargs}                      
\usepackage[colorinlistoftodos,prependcaption, textsize=small]{todonotes}
\newcommandx{\change}[2][1=]{\todo[inline,linecolor=red,backgroundcolor=red!25,bordercolor=red,#1]{#2}}
\newcommandx{\unsure}[2][1=]{\todo[inline,linecolor=blue,backgroundcolor=blue!25,bordercolor=blue,#1]{#2}}
\newcommandx{\info}[2][1=]{\todo[inline,linecolor=green,backgroundcolor=green!25,bordercolor=green,#1]{#2}}
\newcommandx{\improvement}[2][1=]{\todo[inline,linecolor=Plum,backgroundcolor=Plum!25,bordercolor=Plum,#1]{#2}}
\newcommandx{\thiswillnotshow}[2][1=]{\todo[disable,#1]{#2}}

\usepackage{xifthen}
\usepackage[normalem]{ulem}	

\newcounter{cmmnt}
\newcommand{\cmmnt}[4][]{%
	\refstepcounter{cmmnt}%
	{%
		\ifthenelse{\isempty{#1}}{%
			\todo[color=#2!30,linecolor=#2,size=\small]{%
				\sffamily
				\textbf{Comment~\thecmmnt [\uppercase{#3}]:} #4}
		}{%
		\todo[color=#2!30,linecolor=#2,size=\small,#1]{%
			\sffamily
			\textbf{Comment~\thecmmnt [\uppercase{#3}]:} #4}
	}
}}

\usepackage{perpage} 
\MakePerPage{footnote} 

\usepackage{multirow}

\usepackage{algorithm, algpseudocode}

\usepackage[nomain,acronym,toc]{glossaries}
\makeglossaries

\usepackage{tikz}
\usepackage{pgfplots}
\usetikzlibrary{shapes,arrows}
\pgfplotsset{grid style={dashed}}

\usepackage{mathtools}

\usepackage{subcaption}
\copyrightyear{2018} 
\acmYear{2018} 
\setcopyright{acmlicensed}
\acmConference[GECCO '18]{Genetic and Evolutionary Computation Conference}{July 15--19, 2018}{Kyoto, Japan}
\acmPrice{15.00}
\acmDOI{10.1145/3205455.3205457}
\acmISBN{978-1-4503-5618-3/18/07}

\newacronym{NHGRI}{\texttt{NHGRI}} {National Human Genome Research Institute} %

\newacronym{CX}{\texttt{CX}}{Cycle Crossover}
\newacronym{OX}{\texttt{OX}}{Order Crossover}
\newacronym{PMX}{\texttt{PMX}}{Partial Mapped Crossover}

\newacronym{GPU}{\texttt{GPU}}{Graphical Processing Unit}
\newacronym{RNA}{\texttt{RNA}}{Ribonucleic Acid}
\newacronym{OP}{\texttt{OP}}{Optimization Problem}
\newacronym{CS}{\texttt{CS}}{Computer Science}
\newacronym{EA}{\texttt{EA}}{Evolutionary Algorithm}
\newacronym{EAs}{\texttt{EAs}}{Evolutionary Algorithms}
\newacronym{GA}{\texttt{GA}}{Genetic Algorithm}
\newacronym{SOP}{\texttt{SOP}}{Single-Objective Problem}
\newacronym{SOPs}{\texttt{SOPs}}{Single-Objective Problems}
\newacronym{MOO}{\texttt{MOO}}{Multi-Objective Optimization}
\newacronym{MOP}{\texttt{MOP}}{Multi-Objective Optimization Problem}
\newacronym{MOPs}{\texttt{MOPs}}{Multi-Objective Optimization Problems}
\newacronym{MA}{\texttt{MA}}{Memetic Algorithm}
\newacronym{MAs}{\texttt{MAs}}{Memetic Algorithms}
\newacronym{QAP}{\texttt{QAP}}{Quadratic Assignment Problem}
\newacronym{bQAP}{\texttt{bQAP}}{bi-objective Quadratic Assignment Problem}
\newacronym{mQAP}{\texttt{mQAP}}{Multi-Objective Quadratic Assignment Problem}
\newacronym{mQAPs}{\texttt{mQAPs}}{Multi-Objective Quadratic Assignment Problems}
\newacronym{II}{\texttt{II}}{Inverted Index}
\newacronym{IIs}{\texttt{IIs}}{Inverted Indexes}
\newacronym{EMOO}{\texttt{EMOO}}{Evolutionary Multi-Objective Optimization}
\newacronym{MOEA}{\texttt{MOEA}}{Multi-Objective Evolutionary Algorithm}
\newacronym{MOEAs}{\texttt{MOEAs}}{Multi-Objective Evolutionary Algorithms}
\newacronym{MOMA}{\texttt{MOMA}}{Multi-Objective Memetic Algorithm}
\newacronym{MOMAs}{\texttt{MOMAs}}{Multi-Objective Memetic Algorithms}

\newacronym{GD}{\texttt{GD}}{Generational Distance}
\newacronym{MGD}{\texttt{GD$^+$}}{Modified Generational Distance}
\newacronym{IGD}{\texttt{IGD}}{Inverted Generational Distance}
\newacronym{MIGD}{\texttt{IGD$^+$}}{Modified Inverted Generational Distance}
\newacronym{DM}{\texttt{DM}}{Delta Measure}
\newacronym{HV}{\texttt{HV}}{Hypervolume}
\newacronym{IHV}{\texttt{IHV}}{Inverted Hypervolume}
\newacronym{SP}{\texttt{SP}}{Spacing}
\newacronym{DV}{\texttt{DV}}{Diversity}
\newacronym{RC}{\texttt{RC}}{Radial Coverage}
\newacronym{MS}{\texttt{MS}}{Maximum Spread}
\newacronym{ER}{\texttt{ER}}{Error Ratio}
\newacronym{SCC}{\texttt{SCC}}{Success Counting}
\newacronym{CM}{\texttt{CM}}{Coverage Metric}

\newacronym{SLS}{\texttt{SLS}}{Stochastic Local Search}
\newacronym{PLS}{\texttt{PLS}}{Pareto Local Search}
\newacronym{TPLS}{\texttt{TPLS}}{Two-Phase Local Search}
\newacronym{RoTS}{\texttt{RoTS}}{Robust Tabu Search}
\newacronym{BLS}{\texttt{BLS}}{Breakout Local Search}

\newacronym{MOACO}{\texttt{MO-ACO}} {Multi-Objective Ant Colony Optimization} %
\newacronym{SPEATWO}{\texttt{SPEA2}} {Strength Pareto Evolutionary Algorithm II} 
\newacronym{MOEAD}{\texttt{MOEA/D}} {Multiobjective Evolutionary Algorithm Based on Decomposition} %
\newacronym{NSGATWO}{\texttt{NSGA-II}} {Nondominated Sorting Genetic Algorithm II} %

\newacronym{ITIM}{\texttt{ITIM}}{Iterative Improvement}
\newacronym{OAV}{\texttt{OAV}}{Organic Air Vehicles}

\newacronym{MOMGAII}{\texttt{MOMGA-II}} {Multi-Objective Fast Messy Genetic Algorithm}%
\newacronym{MOMGAIIa}{\texttt{MOMGA-IIa}} {Multi-Objective Fast Messy Genetic Algorithm a}%

\newacronym{GRASP}{\texttt{GRASP}}{Greedy Randomized Adaptive Search Procedure}
\newacronym{mGRASP}{\texttt{mGRASP/MH}}{Multi-Objective Greedy Randomized Adaptive Search Procedure}
\newacronym{FDC}{\texttt{FDC}}{Fitness Distance Correlation}
\newacronym{MOGWW}{\texttt{MOGWW}}{Multi-Objective Go with Winners}

\newacronym{GWW}{\texttt{GWW}} {Go With Winner} %
\newacronym{mPLS}{\texttt{mPLS}}{Multi-Objective Pareto Local Search}

\newacronym{AIS}{\texttt{AIS}}{Artificial Immune System}

\newacronym{GISMOO}{\texttt{GISMOO}}{Multi-Objective Genetic Immune System} 
\newacronym{PMSMO}{\texttt{PMSMO}}{PMSMO}

\newacronym{TA}{\texttt{TA}}{Transgenetic Algorithm}
\newacronym{TAs}{\texttt{TAs}}{Transgenetic Algorithms}

\newacronym{NSTA}{\texttt{NSTA}}{Non-Dominated Sorting Transgenetic Algorithms}
\newacronym{MOTAD}{\texttt{MOTA/D}}{Multi-Objective Transgenetic Algorithm/Decomposition}  

\newacronym{GNA}{\texttt{GNA}}{Global Network Alignment}
\newacronym{SIMD}{\texttt{SIMD}}{Single Instruction, Multiple Data}

\newacronym{TSP}{\texttt{TSP}}{Traveling Salesman Problem}
\newacronym{ROE}{\texttt{ROE}}{Reduce-Optimize-Expand}

\newacronym{CIBM}{\texttt{CIBM}}{Centre for Bioinformatics, Biomarker Discovery and Information-Based Medicine}

\newacronym{SDP}{\texttt{SDP}}{Semidefinite Programming}

\newacronym{DMLS}{\texttt{DMLS}}{Dominance Based Local Search}
\newacronym{KNNG}{\textit{k}-\texttt{NNG}}{\textit{k}-Nearest Neighbors Graph}
\newacronym{tSNE}{t-\texttt{SNE}}{t-Distributed Stochastic Neighbor Embedding}
\newacronym{PCA}{\texttt{PCA}}{Principal Component Analysis}
\newacronym{LDA}{\texttt{LDA}}{Linear Discriminant Analysis}
\newacronym{LLE}{\texttt{LLE}}{Local Linear Embedding}
\newacronym{EDA}{\texttt{EDA}}{Exploratory Data Analysis}
\newacronym{RP}{\texttt{RP}}{Random Projection Tree}
\newacronym{RPs}{\texttt{RPs}}{Random Projection Trees}

\begin{document}

\newcommand{\citetemp}[1][missing citation]{(\textcolor{red}{#1})} %
\newcommand{\etal}{et al.} 
\newcommand{\pasmoqap}{\textsc{PasMoQAP}}
\newcommand{\approachname}{mQAPViz}
\newcommand{\myapproach}{\textsc{\approachname}}
\newcommand{\largemyapproach}{\textsc{\large \approachname}}
\newcommand{\hugemyapproach}{\textsc{\huge \approachname}}

\renewcommand{\algorithmicrequire}{\textbf{Input:}}
\renewcommand{\algorithmicensure}{\textbf{Output:}}

\title{\approachname: A divide-and-conquer multi-objective optimization algorithm to compute large data visualizations}

\author{Claudio Sanhueza, Francia Jim{\'e}nez, Regina Berretta, and Pablo Moscato}
\affiliation{
  \institution{School of Electrical Engineering and Computing, The University of Newcastle, NSW, Australia.}
}
\email{{claudio.sanhuezalobos, francia.jimenezfuentes, regina.berretta, pablo.moscato}@newcastle.edu.au}

\renewcommand{\shortauthors}{C. Sanhueza et. al.}
\renewcommand{\shorttitle}{\approachname}

\begin{abstract}
Algorithms for data visualizations are essential tools for transforming data into useful narratives.
Unfortunately, very few visualization algorithms can handle the large datasets of many real-world scenarios.
In this study, we address the visualization of these datasets as a \acrlong{MOP}.
We propose \myapproach, a divide-and-conquer multi-objective optimization algorithm to compute large-scale data visualizations.
Our method employs the \acrfull{mQAP} as the mathematical foundation to solve the visualization task at hand.
The algorithm applies advanced sampling techniques originating from the field of machine learning and efficient data structures to scale to millions of data objects.
The algorithm allocates objects onto a 2D grid layout.
Experimental results on real-world and large datasets demonstrate that \myapproach~is a competitive alternative to existing techniques.
\end{abstract}

\ccsdesc[300]{Applied computing~Multi-criterion optimization and decision-making}
\ccsdesc[300]{Human-centered computing~Visualization}

\keywords{Multi-Objective Optimization, Visualization, Large Datasets}

\maketitle

\section{Introduction}
\label{sec:introduction}
While we read this sentence, terabytes of data have been collectively generated across the globe through many devices we use daily.
\acrfull{EDA} is a critical first stage to investigate datasets.
The techniques that enable \acrshort{EDA} provide a summarized view of a whole dataset.
In particular, visualization algorithms play a relevant role in these tasks.
They may capture some of the inherent hidden characteristics and structures, presenting them in such a way that allow transforming raw data into actionable insights.

Displaying data is a non-trivial task considering that we aim to summarize complex relationships between objects in a human-readable layout (i.e., 2D or 3D).
Visualization techniques are well suited for small (hundreds) and medium (thousands) size datasets.
However, the task is even more challenging today considering the scale of the datasets in need of these algorithms. 
Data products have been an instrumental part of big technological companies. 
Large institutions have the human capital and the computational infrastructures to conduct big analyses on massively distributed systems.
However, there are many researchers and practitioners who do not have access to these platforms. 
They would benefit from a new generation of more efficient algorithms that can compute visualizations of large datasets on modern multi-core workstation computers.

Typically, \acrshort{EDA} is used to collect new insights from independent views obtained from the data.
We argue however that algorithms devoted to data analysis should take into account several viewpoints to evaluate relations.
Multi-objective optimization algorithms have been widely proposed and used to address real-world \acrfull{MOPs} \cite{deb2001multi}.
In \acrshort{MOPs}, the challenge is to simultaneously satisfy \emph{multiple} and possibly \emph{conflicting} objectives.

In multi-objective optimization problems, there is not a unique solution, but a set of non-dominated solutions (i.e., a trade-off in the objective space) which is called the Pareto optimal set.
\acrfull{MOEAs} are well-suited to approximate the Pareto optimal set in a broad variety of \acrshort{MOPs} \cite{zitzler1999multiobjective}.
\acrshort{MOEAs} evolve individuals (solutions) typically organized in populations, exploring the solution space using operators such as recombination, mutation, and selection to improve the population.

In this study, we propose \myapproach~(pronounced \emph{mapviz}), a novel multi-objective optimization algorithm to compute visualizations of large-scale datasets.
Our algorithm employs the \acrlong{mQAP}, which is presented in the following sections, as the mathematical model to position objects in a grid layout.
To the best of our knowledge, this study is the first one that addresses the visualization of large datasets as a \acrlong{MOP}.
In particular, we present the following contributions:

\begin{itemize}
\item{We propose \myapproach, a new multi-objective optimization algorithm which can compute large-scale data visualizations.}
\item{We propose a divide-and-conquer approach to solve sub-problems of the \acrlong{mQAP} to tackle the visualization task at hand.}
\item{We evaluate our approach on a set of large and real datasets that belong to different domains against the state-of-the-art visualization algorithm \acrshort{tSNE}.}
\end{itemize}

We organize the remainder of the paper as follows. In Section~\ref{sec:related_work}, we discuss the related work.
We present the details of \myapproach~in Section~\ref{sec:methodology}. We discuss our experimental methodology and the results in Section~\ref{sec:experiments}.
We finish with the conclusions and future work in Section \ref{sec:conclusions}.

\section{Related Work}
\label{sec:related_work}
Data visualizations assist the process of representing data for supporting the tasks of exploration, confirmation, presentation, and understanding to deliver knowledge \cite{few2009now}. 
Several tools and algorithms have been developed over the years.
For example, force-directed layout algorithms use a graph data structure to model datasets as a dynamical system.
Nodes represent mutually repelling particles, and edges correspond to the existence of an attractive force between them.
The layout is determined once the forces drive the system to equilibrium \cite{fruchterman1991graph}.

Other approaches organize objects in a grid layout.
These methods produce a visualization using a finite number of positions defined by a grid.
For example, the grid layout has been used to visualize biochemical networks~\cite{li2005grid}.
Another general method for data visualization using a grid layout is presented in \cite{abbiw2006divide} in which the authors proposed a divide-and-conquer method that recursively distributes the data in grids.
Later, \textsc{QAPgrid} was proposed in~\cite{inostroza2011qapgrid} which using the \acrfull{QAP} as the mathematical model, a proximity graph, and a single-objective optimization guides the generation of a grid layout of objects.
The method has been used in several applications~\cite{clark2012genome, warren2017using}, but its current version cannot compute large datasets (i.e., millions of objects) in a reasonable time.

Data visualization can also be seen as the task of mapping data from a high-dimensional to a low-dimensional space using some distance-preserving dimensionality reduction in the final representation. 
Traditional methods in the literature include \acrfull{PCA} or \acrfull{LDA}.
Other well-known algorithms use the hypothesis that the data can be approximated by a low-dimensional manifold, such as Laplacian Eigenmaps~\cite{belkin2003laplacian}, Isomap~\cite{tenenbaum2000global} or 
\acrfull{LLE}~\cite{roweis2000nonlinear}.
We refer to Ref.~\cite{van2009dimensionality} in which the authors presented a comparative study of dimensionality reduction techniques.

A handful of visualization algorithms can efficiently address the challenges of large-scale datasets.
In particular, two successful algorithms \cite{maaten2014accelerating, tang2016visualizing} that aim at closing this gap share two characteristics: efficient data structures and ad-hoc probabilistic models.
The popular \acrshort{tSNE} method minimizes the divergence between a distribution that measures pairwise similarities of the input objects
and a distribution that measures pairwise similarities of the corresponding low-dimensional representation.
\textsc{LargeVis} implements an approximate $k$-nearest neighbor graph and graph sampling techniques, improving the original complexity from $\mathcal{O}(n~log~n)$ to $\mathcal{O}(n)$ (in which $n$ is the number of samples in the dataset).
In the following sections, we present \myapproach~which integrates ideas taken from the multi-objective optimization domain to compute large data visualizations.

\section{A multi-objective optimization algorithm to compute large data visualizations}
\label{sec:methodology}
\begin{figure*}[ht]
	\centering
	\includegraphics[width=0.75\textwidth]{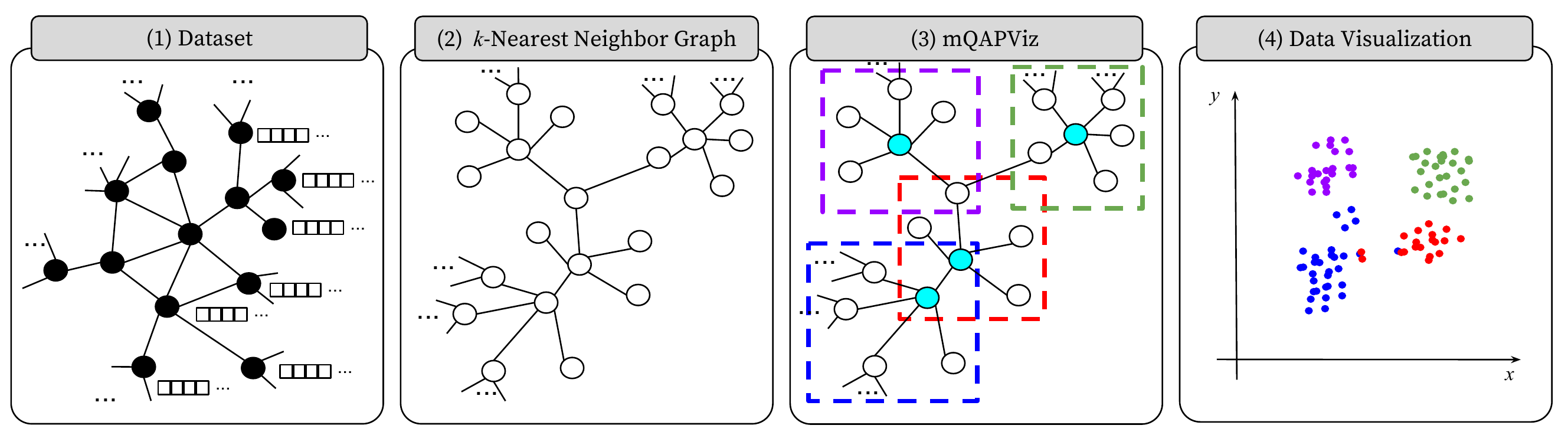}
	\caption{An illustration of the workflow involved in \myapproach.}
	\label{methodology:fig:general_pipeline}
\end{figure*}

In this section we present \myapproach, a new multi-objective algorithm to generate visualizations of large-scale datasets.

\subsection{The Multi-Objective Quadratic Assignment Problem approach data visualization} 

Formally, the special case of the \acrfull{mQAP} in which the number of objects is the same to the number of positions ($n$) is defined as follows:

\begin{equation}
\label{eq:formal_mqap}
\begin{aligned}
	\underset{\pi \in P_n}{\text{minimize}}\;\;
			C(\pi)	 & = \{C_1(\pi), C_2(\pi), ... , C_k(\pi) \} \\
			C_r(\pi)	 & = \sum_{i=1}^{n} \sum_{j=1}^{n} d_{ij}f^r_{\pi(i)\pi(j)}, \: r=1,\dots,k~,
\end{aligned}
\end{equation}

\noindent
where $f^r_{\pi(i)\pi(j)}$ represents the flow between the object $\pi(i)=p$ and $\pi(j)=q$ of the $r$-th flow and $d_{ij}$ is the distance between the position $i$ and $j$. $P_n$ represents the set of all permutations $\pi: N \rightarrow N$.
The product $d_{ij}\,f^r_{\pi(i)\pi(j)}$ corresponds to the $r$-th cost of allocating object $\pi(i)=p$ to the position $i$ and object $\pi(j)=q$ to the position $j$.
The difference between \acrshort{mQAP} and the single-objective \acrfull{QAP} is that we consider more than one flow (i.e., $k$ flows in \autoref{eq:formal_mqap}), and we minimize them simultaneously.

Using the \acrshort{QAP} as a proxy for data visualization is a simple and intuitive idea.
First, we create a layout with available positions to allocate the objects.
We can create a human-readable layout to visualize a dataset with $n$ objects using a low-dimensional grid with $m$ possible positions ($n << m$).
Second, we allocate the objects into the layout with the aim of minimizing the cost which is a function of the distances and the flows.
Intuitively, \emph{similar} objects should be positioned closer to each other and \emph{dissimilar} objects should be pushed away. We may define a (dis)similarity measure depending on the particular domain of study.
Although this approach has shown relatively good results in datasets with thousands of objects~\cite{inostroza2011qapgrid}, it is impractical for large-scale datasets, and our contribution is addressing this need.

\subsection{\myapproach}
\label{subsec:mqapviz}
To compute visualizations of large datasets, we use a divide-and-conquer strategy which creates and solves several \acrshort{mQAP} instances.
These sub-instances represent a sampled portion of the whole dataset.
Our method is based on two main components. 
An initial layout is created, and later it is optimized thanks to a \acrshort{mQAP}-based approach.
For this second part, we compute a \acrfull{KNNG} which is used to obtain information about the most similar sets of the data objects.
Then, a sampling strategy is used to select a set of nodes that will be used to create \acrshort{mQAP} sub-instances to be optimized.
Each sub-instance is optimized thanks to our method called \pasmoqap~\cite{sanhueza2017pasmoqap}, a parallel asynchronous memetic algorithm.
Finally, we merge the individual solutions to create several visualizations in a low-dimensional space.

The \autoref{methodology:fig:general_pipeline} illustrates the workflow involved in \myapproach~(\autoref{methodology:algo:mqapvis}).
In the next sections, we discuss the details of the main components implemented in \myapproach. 

\begin{algorithm}[t]
	\small
    \caption{Pseudo-code of \myapproach}
    \label{methodology:algo:mqapvis}
    
    \begin{algorithmic}[1] 
	    \Require dataset $D$, number of trees $n_t$, number of neighbors $k$, number of iterations $it$, portion of sampled nodes $p_s$
	    \Ensure Final set of visualizations $L_f$
		    \State $\mathcal{P}^{*} \gets \emptyset$ \Comment{Set of individual solutions}
		    \State $G_{k-nn} \gets \Call{BuildKNNG}{D, n_t, k, it}$ \Comment{See \autoref{methodology:algo:knn_construction}}
		    \State $L_0 \gets \Call{LargeVis}{D, G_{k-nn}}$
	        \State $V_s \gets \Call{NegativeSampling}{G_{k-nn}, p_s}$
	        
	        \For{each vertex $v_i \in V_s$ \textbf{in parallel}} \Comment{Solving \acrshort{mQAP} instances}
		        \State $I_{v_i} \gets \Call{CreateMQAPVisInstance}{G_{k-nn}, v_i}$
		        \State $\mathcal{P}^{*}_{v_i} \gets \Call{PasMoQAP}{I_{v_i}}$
		        \State $\mathcal{P}^{*} \gets \mathcal{P}^{*} \cup \mathcal{P}^{*}_{v_i}$
	        \EndFor

			\State $V_s \gets \emptyset$
			\State $V_s \gets \Call{GetMissingVertices}{\mathcal{P}^{*}}$ \Comment{Getting vertices without an assigned position}
	        \For{each vertex $v_i \in V_s$ \textbf{in parallel}}
		        \State $I_{v_i} \gets \Call{CreateMQAPInstance}{G_{k-nn}, v_i}$
		        \State $\mathcal{P}^{*}_{v_i} \gets \Call{PasMoQAP}{I_{v_i}}$
		        \State $\mathcal{P}^{*} \gets \mathcal{P}^{*} \cup \mathcal{P}^{*}_{v_i}$
	        \EndFor
	
			\State $L_f \gets \Call{MergeSolutions}{\mathcal{P}^{*}}$

            \State \textbf{return} $L_f$ \Comment{Return the visualizations}
    \end{algorithmic}
\end{algorithm}

\subsection{Building the \emph{k}-Nearest Neighbors Graph}
The \acrfull{KNNG} is a critical data structure in modern machine learning applications.
Formally, a \acrshort{KNNG} can be defined as a graph $G_{k-nn} = (V,E)$ where $V=\{v_1, ..., v_n\}$ is the vertex set and $E=\{e_1, ..., e_k\}$ is the edge set.
An edge $e_i = (v_p, v_q) \in E$, $v_p$, $v_q \in V$, exists if $v_p \ne v_q$ and if either $v_q$ is one of $k$-nearest neighbors of $v_p$ (or viceversa, or both) under a particular similarity measure. 
The computation of a $G_{k-nn}$ has a time complexity of $\mathcal{O}(n^2)$ which is impractical on large datasets.
In our approach, we compute the $G_{k-nn}$ using an algorithm that iteratively refines an initial $G_{k-nn}$ approximation.
The simple idea behind this method is that \emph{``the neighbor of my neighbor is probably my neighbor.''}
This algorithm outperforms previous approaches in efficiency and accuracy \cite{tang2016visualizing}.

The \textsc{BuildKNNG} algorithm (\autoref{methodology:algo:knn_construction}) begins creating $n_t$ \acrfull{RPs}.
A \acrlong{RP}, which is a variant of the k-d tree spatial data structure \cite{bentley1975multidimensional}, automatically adapts to an intrinsic low dimensional structure.
Using the \acrshort{RPs}, the next step is to find the $k$-nearest neighbors of each data sample (line 3, \autoref{methodology:algo:knn_construction}).
These neighbors can be used as the initial $G_{k-nn}$ which the algorithm refines using an iterative procedure (lines 5 to 21, \autoref{methodology:algo:knn_construction}).
Later, for each data object $s_i$ and each neighbor $s_j \in knn(s_i)$ the algorithm includes the distance between $s_i$ and the neighbors of $s_j$, i.e., $s_p \in knn(s_j)$, in a heap data structure $H(s_i)$.
The algorithm updates the $k$-nearest neighbors of each sample $s_i$ using $H(s_i)$ (line 19, \autoref{methodology:algo:knn_construction}).
The algorithm creates the final $G_{k-nn}$ using the neighbors computed during the iterative refinement process.

\begin{algorithm}[t]
    \small
    \caption{Pseudo-code of \textsc{BuildKNNG}}
    \label{methodology:algo:knn_construction}
    
    \begin{algorithmic}[1] 
	    \Require dataset $D$, number of trees $n_t$, number of neighbors $k$, number of iterations $it$
	    \Ensure k-Nearest Neighbor Graph $G_{k-nn}$
		    \State $RPT \gets \Call{GetRandomProjectionTrees}{D, n_t}$
	        \For{each sample $s_i \in D $ in parallel}
		        \State $knn(s_i) \gets \Call{SearchNN}{RPT,s_i,k}$ \Comment{Search in the random projection trees the $s_i$'s $k$ nearest neighbors}
	        \EndFor

			\For{$t = 1$ to $it$}
				\State $knn_{old}()= knn()$
				\State $knn() \gets \emptyset$
		        \For{each sample $s_i \in D$ in parallel}
			        \State $H(s_i) \gets \Call{CreateMaxHeap}$
			        \For{$s_j \in knn_{old}(s_i)$}
				        \For{$s_p \in knn_{old}(s_j)$}
					        \State $d_{s_i,s_p} \gets dist(s_i, s_p)$ \Comment {$|| \overrightarrow{s_i} - \overrightarrow{s_p}||$}
					        \State $H(s_i).push(s_p, d_{s_i,s_p})$
					        \If{$H(s_i).size() > k$}
						        \State $H(s_i).pop()$
					        \EndIf
					    \EndFor
			        \EndFor
			        \State $knn(s_i) \gets H(s_i)$
		        \EndFor
		    \EndFor
		     \State {$G_{k-nn} \gets \emptyset$}
		    \For{each sample $s_i$ and each $s_j \in knn(s_i)$}
			    \State {$G_{k-nn} \gets \Call{NewEdge}{s_i,s_j}$}
		    \EndFor		

            \State \textbf{return} $G_{k-nn}$
    \end{algorithmic}
\end{algorithm}

\subsection{Creating the mQAP sub-instances via a negative sampling method}
\label{subsec:building_mqap_subinstances_negative_sampling}
To improve the efficiency of \myapproach, we compute the visualizations using a sampled portion of the vertices in $G_{k-nn}$ using the method called \emph{negative sampling}. 
Negative sampling is an alternative method to reduce the computational complexity of model optimization~\cite{mikolov2013distributed}.
Given an initial layout, we then generate several \acrshort{mQAP} sub-instances which \myapproach~solves in parallel.
To generate mQAP sub-instances, we will sample a portion of the total number of data objects using this sampling method.
Negative sampling is widely used in language modeling, and later it was employed in representation learning techniques~\cite{tang2015line,grover2016node2vec}.
We need to take this approach to improve the quality of the visualizations of our initial layout computed by the algorithm called \textsc{LargeVis} \cite{tang2016visualizing} which generates a human-readable layout of large datasets initially described in a high-dimensional space.
\textsc{LargeVis} uses the probabilistic modeling ideas of \acrshort{tSNE} \cite{maaten2008visualizing,maaten2014accelerating}, which has been widely adopted to compute visualizations of high-dimensional data. 

The method is based on sampling multiple \emph{negative edges} (i.e., edges that do not exist in the $G_{k-nn}$) according to some probability distribution for each edge.
Given a vertex $v_p \in G_{k-nn}$, we create a \acrshort{mQAP} sub-instance using $v_p$ and its neighbors $knn(v_p)$.
More explicitly, for each vertex $v_i \in G_{k-nn}$, we randomly sample vertices $v_j \in G_{k-nn}$ according to the probability  distribution that depends on the node degree (i.e., $P_n(v_j) \approx deg(v_j)^{\frac{3}{4}}$, \cite{mikolov2013distributed}, in which $deg(v_j)$ is the degree of vertex $v_j$).
The sampling method thus reduces the number of \acrshort{mQAP} instances that need to be optimized. 
Once the method computes the set of sub-instances, \myapproach~proceeds to optimize them in parallel using \pasmoqap~implementing our divide-and-conquer strategy.

\begin{figure}[t]
	\centering
	\includegraphics[width=0.3\textwidth]{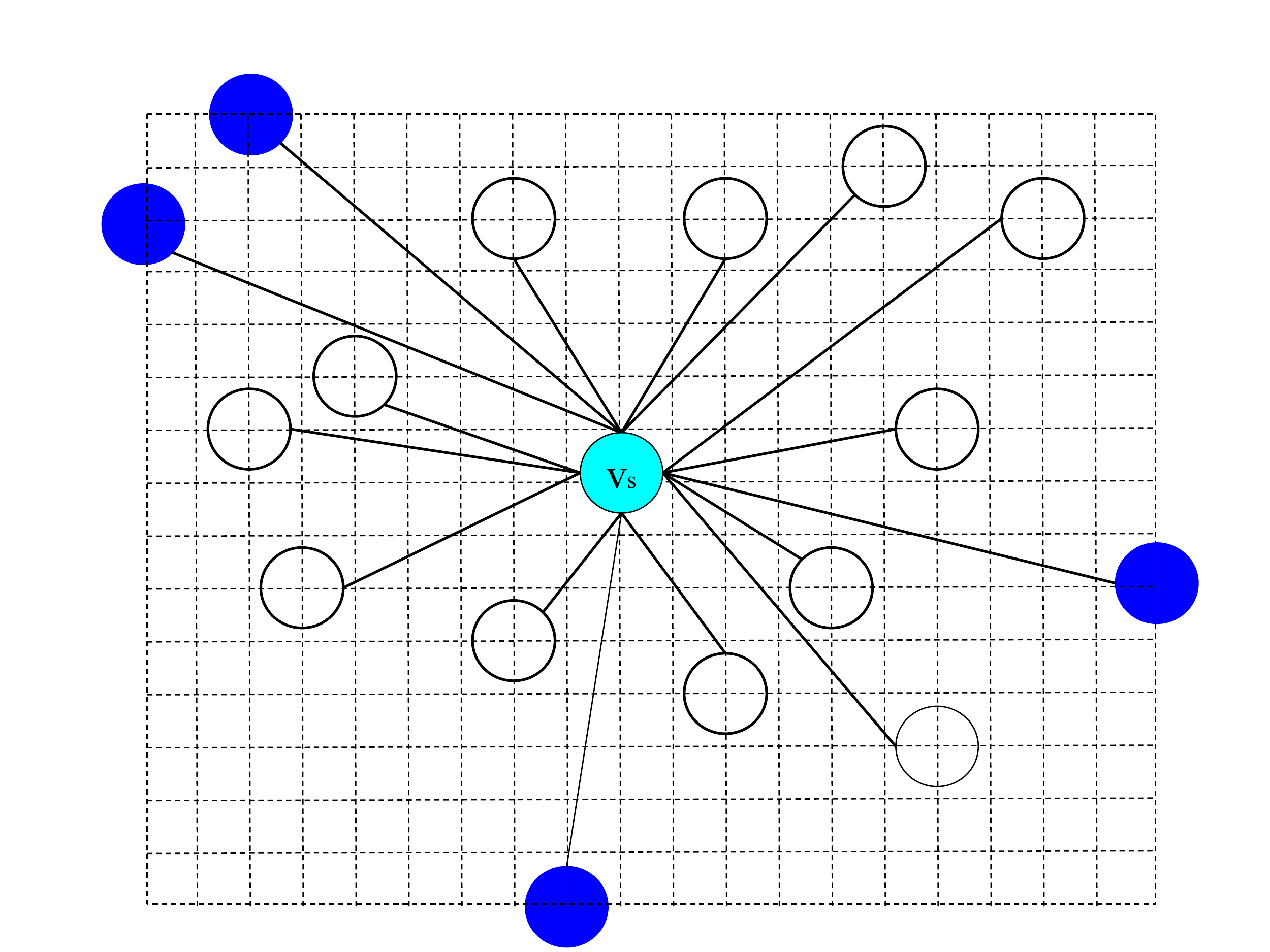}
	\caption{A representation of a \acrshort{mQAP} sub-instance. The vertex $v_s$ and its $k$-nearest neighbors $knn(v_s)$ induce a layout with available positions for the instance. The blue vertices are the lower and upper bounds of the induced layout.}
	\label{methodology:fig:example_induced_layout}
\end{figure}

\subsection{Building mQAP instances for visualization}
\label{subsec:building_mqap_instances_visualization}
In this section, we propose a general method to compute flows and distances for the creation of the \acrshort{mQAP} sub-instances to be optimized.
However, we note that there is not a unique definition of flow between objects and distances between locations, so here we will present our choices for the visualization task given an initial layout of reference.

\textbf{Creating sub-layouts -- } 
Given a sampled vertex $v_s \in G_{k-nn}$, let $p_0(v_s)$ be its initial position in the layout produced by \textsc{LargeVis}. 
We denote the layout induced by a sampled vertex $v_s$ as $L(v_s)$. 
Let $knn(v_s)$ be its $k$-nearest neighbors.
We generate each \acrshort{mQAP} instance as follows.
We first identify the initial positions where the neighbors are located in the layout (to find both the upper and lower bounds of the two coordinates of this group of objects).
These correspond to the blue objects in \autoref{methodology:fig:example_induced_layout}. 
Once the method identifies the bounds of this enclosing rectangle, we need to define the number of grid positions needed.
We have chosen it to be a $50 \times 50$ rectangular grid whose positions are separated by an equal distance based on the computed bounds (assuming that 2500 grid positions allow all objects to be allocated to a grid point).

\textbf{Objective functions -- } 
A \acrshort{mQAP} instance requires the definition of at least two types of flows between the data objects.
In this work, we define two types of flows, the general structure of our approach would eventually handle other heuristic decisions. 

\begin{figure}[t]
	\centering
	\includegraphics[width=0.35\textwidth]{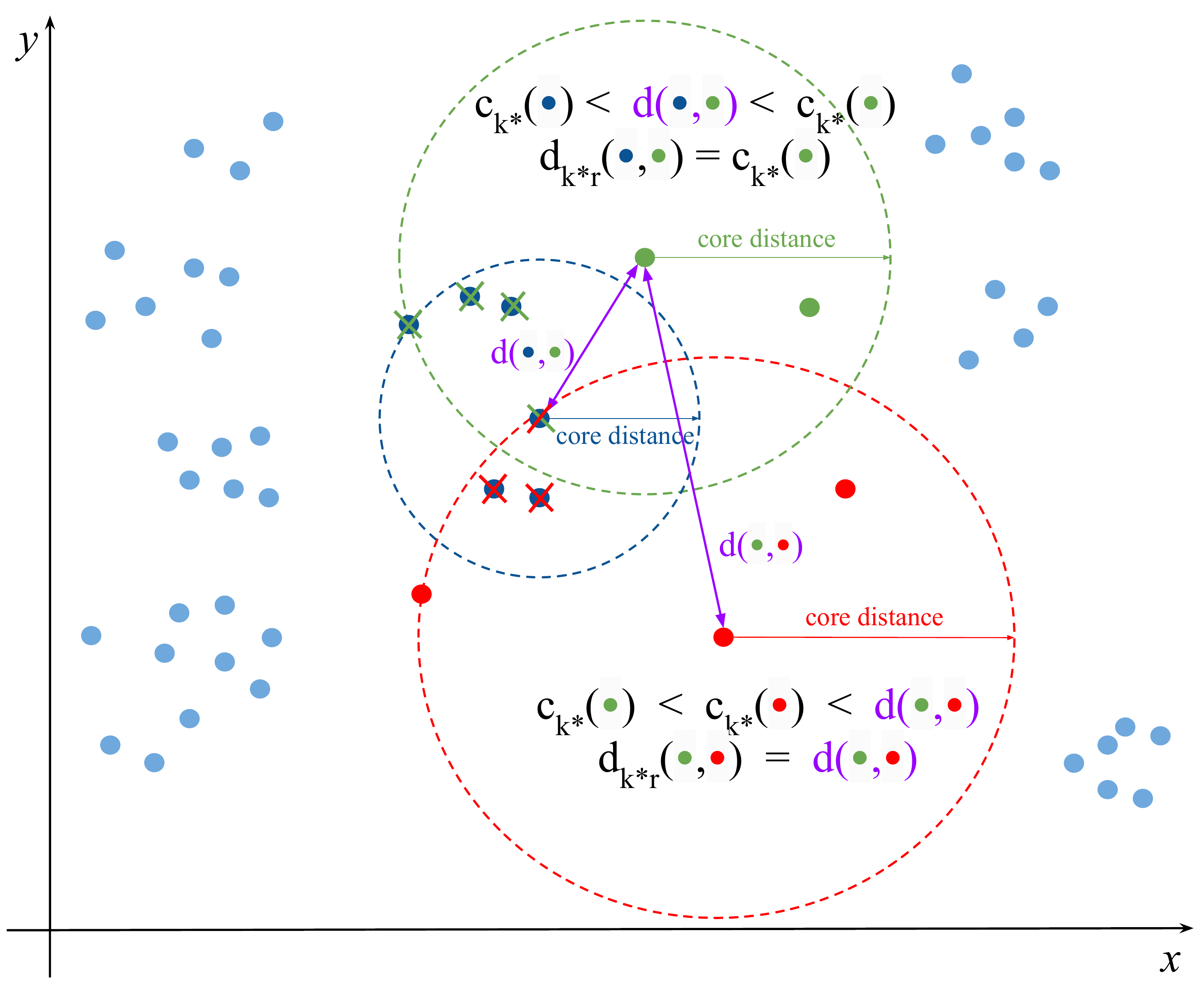}
	\caption{An example of mutual reachability distance computation with $k^*=5$.}
	\label{methodology:fig:mutual_reachability_distance}
\end{figure}

The first definition of flows is motivated to deal with a problem that arises when working with datasets that might contain outliers, e.g., due to the corruption of the values or incorrect measurements.
Intuitively, we want our visualization algorithms to lay objects organized as ``islands'', groups of highly similar objects packed in nearby positions. 
Towards that end, we use a straightforward and low-cost estimation of density centered in a vertex in a layout.
The \emph{$k^*$-core distance} of a object $v$ in a layout, denoted as $c_{k^*}(v)$, corresponds to the Euclidean distance, in the layout, between $v$ and its $k^*$-th nearest neighbor.
We also define the \emph{mutual reachability distance}~\cite{ester1996density, campello2013density, campello2015hierarchical} between two vertices $v_p$, $v_q \in G_{k-nn}$ with parameter $k^*$ as:
\begin{equation}
\begin{aligned}
d_{k^*r}(v_p, v_q) = max \big\{c_{k^*}(v_p),~~c_{k^*}(v_q),~~d(p(v_p), p(v_q))\big\}.
\end{aligned}
\end{equation} 
\autoref{methodology:fig:mutual_reachability_distance} shows how the mutual reachability distance is computed with three objects when $k^*=5$.
First, for the blue object (located near the center of the figure) a circle encloses all objects which are its first five nearest neighbors.
The same is the case for the larger green circle near the top with a different center and the red one near the bottom. 
Thus, the mutual reachability distance between blue and green is equal to the core distance of the green object.
On the other hand, the mutual reachability distance from red to green corresponds to the distance from the center of the red circle to the center of the green one, since it is larger than both core distance.

Then, we define the first of the two flows for our \acrshort{mQAP} instances as follows:
\begin{equation}
\begin{aligned}
\{f^{(1)}_{v_p v_q}\} & = \Big\{\frac{1}{1 + \exp(d_{k^*r}(v_p, v_q))}\Big\},
\end{aligned}
\end{equation} 

\noindent
where $d_{k^*r}(v_p, v_q)$ corresponds to the mutual reachability distance between $v_p$ and $v_q$.

For the second set of flows, we need some further definitions.
Let $v_p$ and $v_q$ be two vertices belonging to the sub-layout $L(v_s)$ and $d(p(v_p), p(v_q))$ the Euclidean distance between the positions assigned to these vertices in a layout. Let $p_0(v_p)$ and $p_0(v_q)$ be the positions assigned in the original layout provided by  \textsc{LargeVis}. Let $d_1=d(p_0(v_p), p_t(v_p))$ and 
$d_2=d(p_0(v_q), p_t(v_q))$ correspond to the Euclidean distance between the initial position assigned to object $v_p$ and the one in which the same object is assigned after iteration $t$ of \pasmoqap.
Also, $d_x = d(p_0(v_p), p_0(v_q))$ is the original Euclidean distance between $v_p$ and $v_q$ in the layout produced by \textsc{LargeVis} and $d_y = d_t(p(v_p), p_t(v_q))$ is their Euclidean distance on a layout $L_t$.
We compute the second type of flow as follows:
\begin{equation}
\begin{aligned}
\{f^{(2)}_{v_p v_q}\} & = \Big\{\frac{1}{(1 + d(v_p,v_q))}\Big\},
\end{aligned}
\end{equation}
\noindent
where $d(v_p,v_q)$ is the arithmetic average of these four distances $d(v_p,v_q)= (d_1 + d_2 + d_x + d_y) / 4$.
We note, however, that this decision is non-standard as the flows depend on the actual position of the objects in a layout (i.e., in some sense ``dynamically'' changing during the optimization process).
We expect that other flow definitions can also be explored in future contributions. 

\textbf{Positioning all the objects in the layout -- }
\myapproach~keeps track of all the objects that have been allocated during the optimization process.
However, it can happen that no all the objects are positioned after the optimization procedure.
This case can happen because our algorithm uses the negative sampling method, selecting vertices and their $k$-nearest neighbors that will be allocated by the algorithm.
In the best case scenario, the sampling technique will select all the objects in the dataset, but the technique cannot ensure it.
In this case, for each not positioned vertex, our algorithm executes the optimization procedure (line 11 to 16 in \autoref{methodology:algo:mqapvis}).
In this way, we guarantee that all the objects are allocated via \myapproach~before generating the final visualizations.

\subsection{Merging solutions from the Pareto fronts}
\label{subsec:merging_solutions}
\begin{figure}[t]
	\centering
	\includegraphics[width=0.33\textwidth]{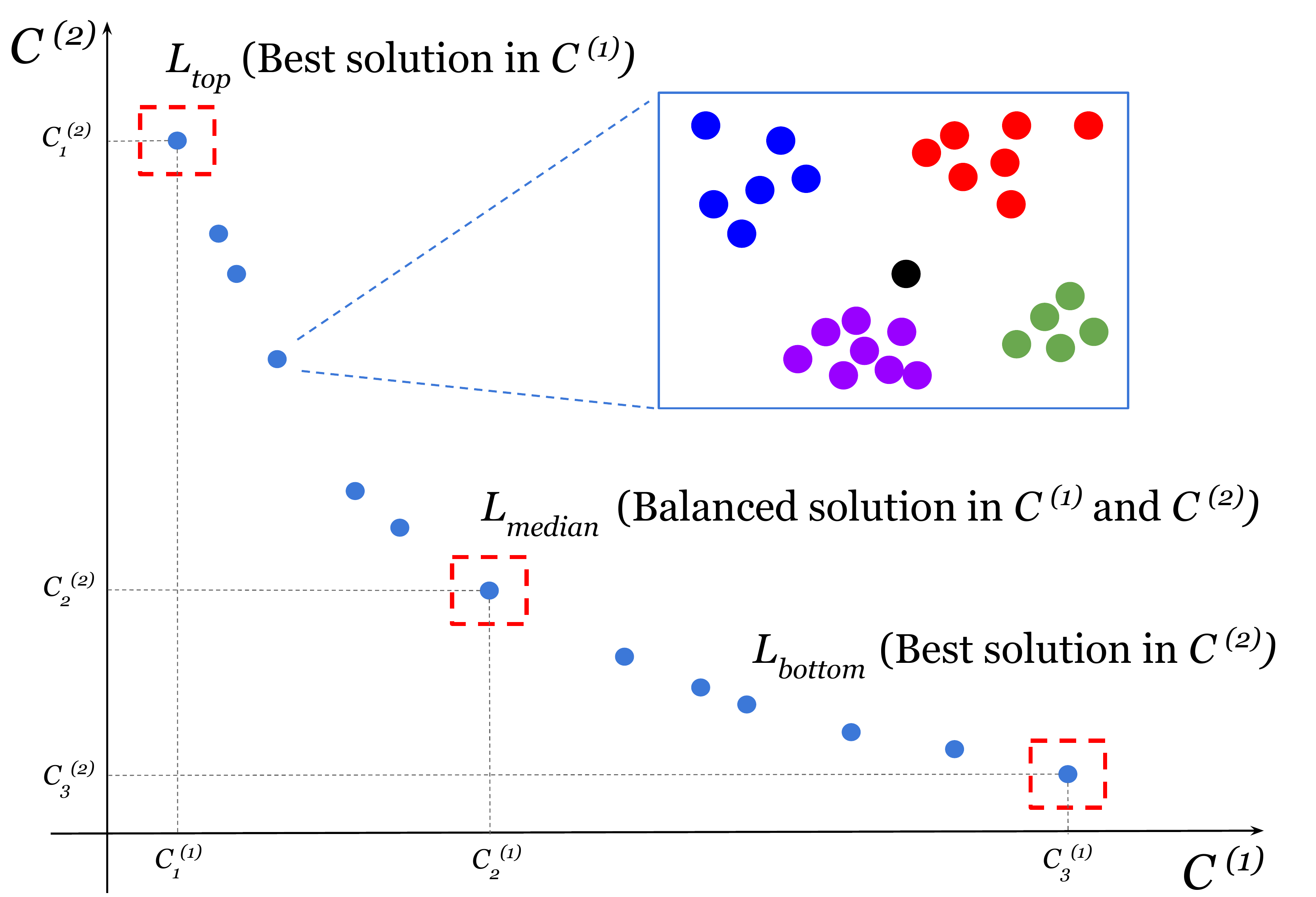}
	\caption{A representation of selecting of solutions to create the final visualizations. Our heuristic selects three visualizations (enclosed by squares) that belong to each Pareto front. The method chooses the best solution according to the cost $C^{(1)}$ ($L_{top}$), the best solution according to the cost $C^{(2)}$ ($L_{bottom}$), and a third solution that balances both objectives ($L_{median}$).}
	\label{methodology:fig:merging_mqapviz_solutions}
\end{figure}

\myapproach~computes many Pareto fronts, one for each vertex that was sampled using the negative sampling procedure. Each front contains several non-dominated solutions.
Consequently, it is frequent that \myapproach~can assigns the same vertex to several available positions in the different layouts.
However, in multi-objective optimization typically a user is interested in only a handful of solutions (from the front), which in our case correspond to different visualizations.
We implement a simple method which merges solutions taken from the Pareto fronts.
Thus, for each Pareto front, our heuristic procedure selects only three solutions (i.e., layouts) that are used to generate the final visualizations.
Let $C^{(1)}$ and $C^{(2)}$ be the cost function which are computed using the flow definitions $f^{(1)}$ and $f^{(2)}$ respectively.
Since we defined a bi-objective optimization problem, we can easily select two extreme solutions in a Pareto front according to the objectives $C^{(1)}$ and $C^{(2)}$ (\autoref{methodology:fig:merging_mqapviz_solutions}).
We called these two solutions $L_{top}$ and $L_{bottom}$, and they represent the best solutions for objectives $C^{(1)}$ and $C^{(2)}$ respectively.
The heuristic selects a third solution, which we called $L_{median}$,  the one closest to the median ranked position of the solutions in the Pareto front after ordering.
$L_{median}$ represents a ``balanced'' trade-off between both objectives functions.
Then, for each computed Pareto front of the same type (e.g `top', `bottom' or `median') (for each sampled vertex), and for each vertex that is not a seed vertex, we allocate it to one not yet allocated position (but assigned in at least one of these Pareto fronts). 
Due to the space restrictions, we only report the resulting visualizations obtained by merging by this process the $L_{median}$ layouts of each Pareto front.

\section{Experiments}
\label{sec:experiments}
In this section, we evaluate \myapproach~quantitatively and qualitatively on several real-life and large datasets.
We implemented \myapproach~in C++ using the framework ParadisEO \cite{liefooghe2010paradiseo}.
We performed the experiments on individual machines in The University of Newcastle's Research Compute Grid that contains a cluster of 32 nodes Intel\textregistered~Xeon\textregistered~CPU E5-2698 v3 @ 2.30 GHz x 32 with 128 GB of RAM.

\begin{table}[ht]
\centering
\caption{A summary of the datasets' statistics.}
\label{tab:experiments:datasets}
\begin{tabular}{lccc}
\toprule
\textbf{Dataset} & \textbf{\# samples} & \textbf{\# dimensions} & \textbf{\# classes} \\
\midrule
Astroph       & 18,772               & 128                   & -                     \\
Pubmed           & 19,717               & 128                   & 3                   \\
MNIST            & 70,000               & 784                   & 10                  \\
Fashion-MNIST    & 70,000               & 784                   & 10                  \\
Flickr           & 80,513               & 128                   & 195                 \\
Pokec            & 1,632,803             & 128                   & -                  \\
Spammers         & 5,321,961             & 128                   & 2                  \\
\bottomrule
\end{tabular}
\end{table}

 \begin{figure*}[ht]
 \centering
 \begin{tabular}{ccc}
   
   \includegraphics[width=0.3\linewidth]{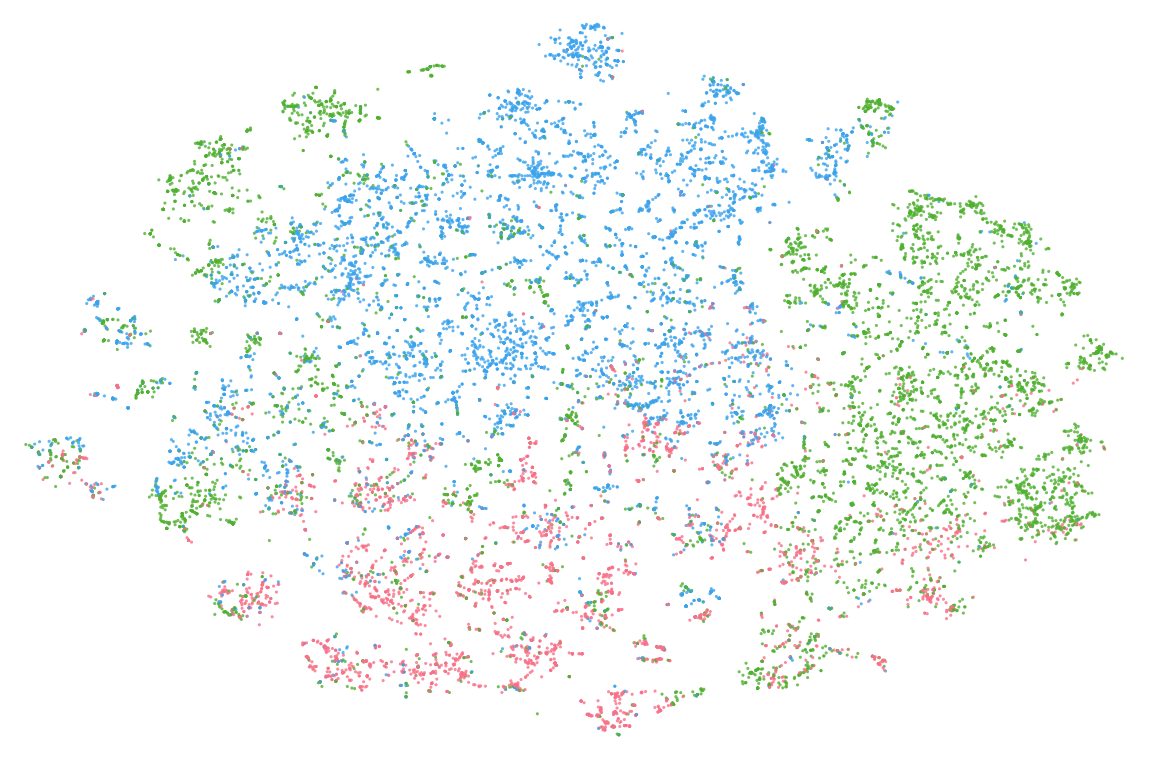} &   
   \includegraphics[width=0.3\linewidth]{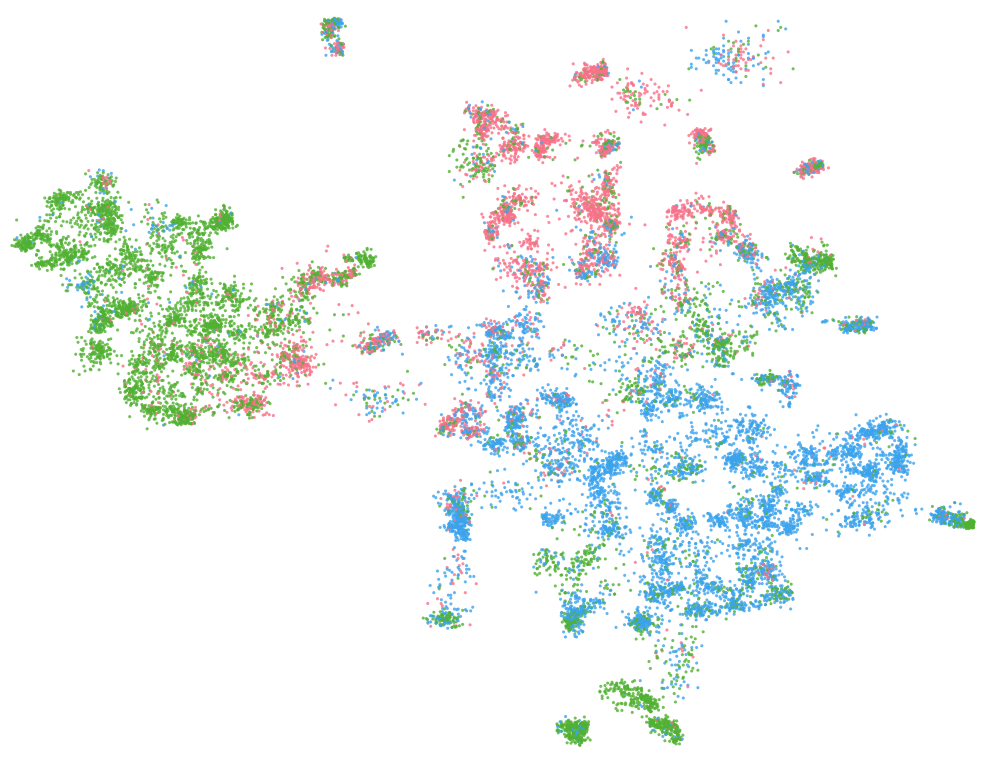} &
   \includegraphics[width=0.3\linewidth]{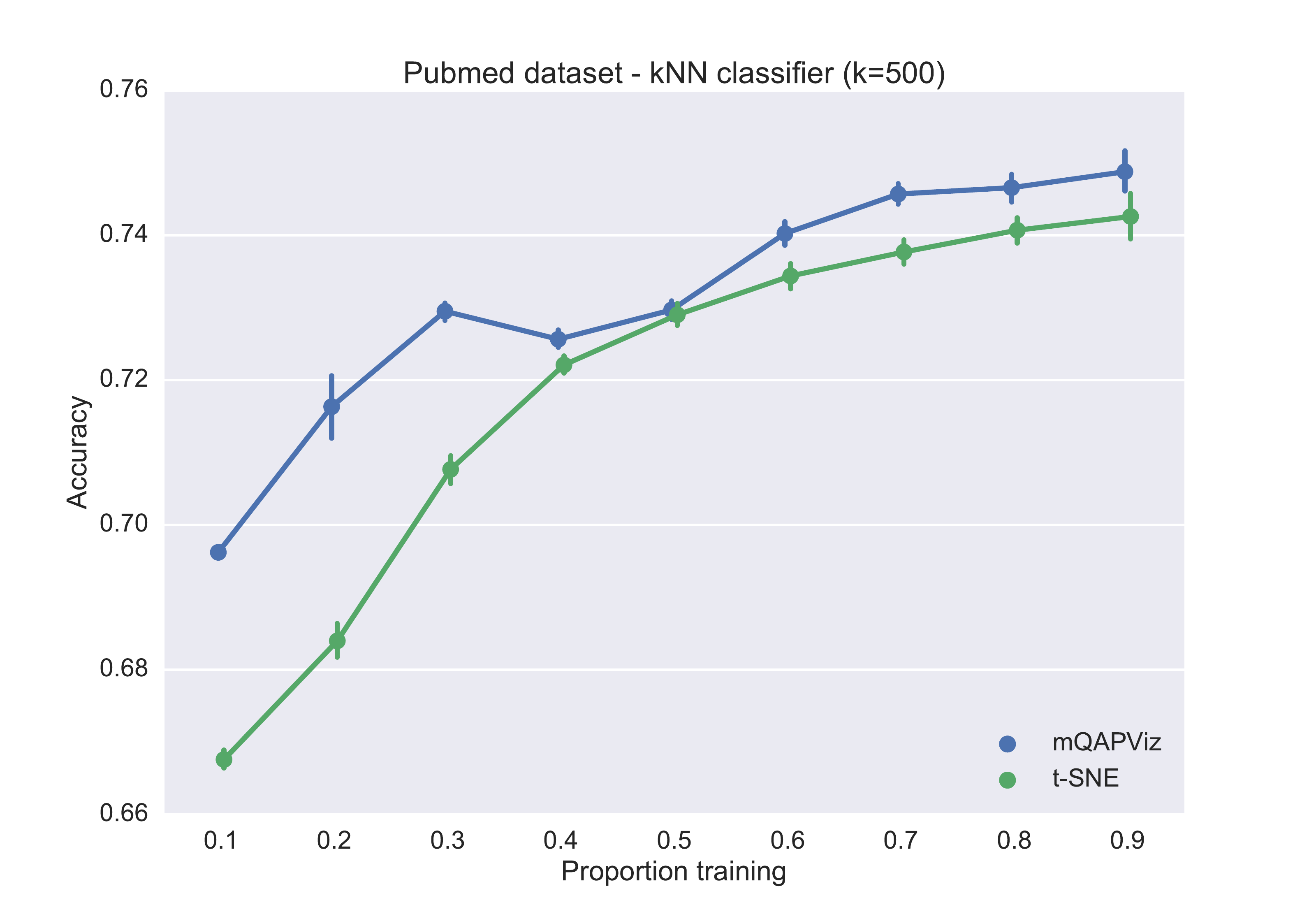} \\ 
   (a) Pubmed (\acrshort{tSNE}) & (b) Pubmed (\myapproach) & (c) Pubmed (\acrshort{tSNE} vs. \myapproach)\\[6pt] 
   \includegraphics[width=0.3\linewidth]{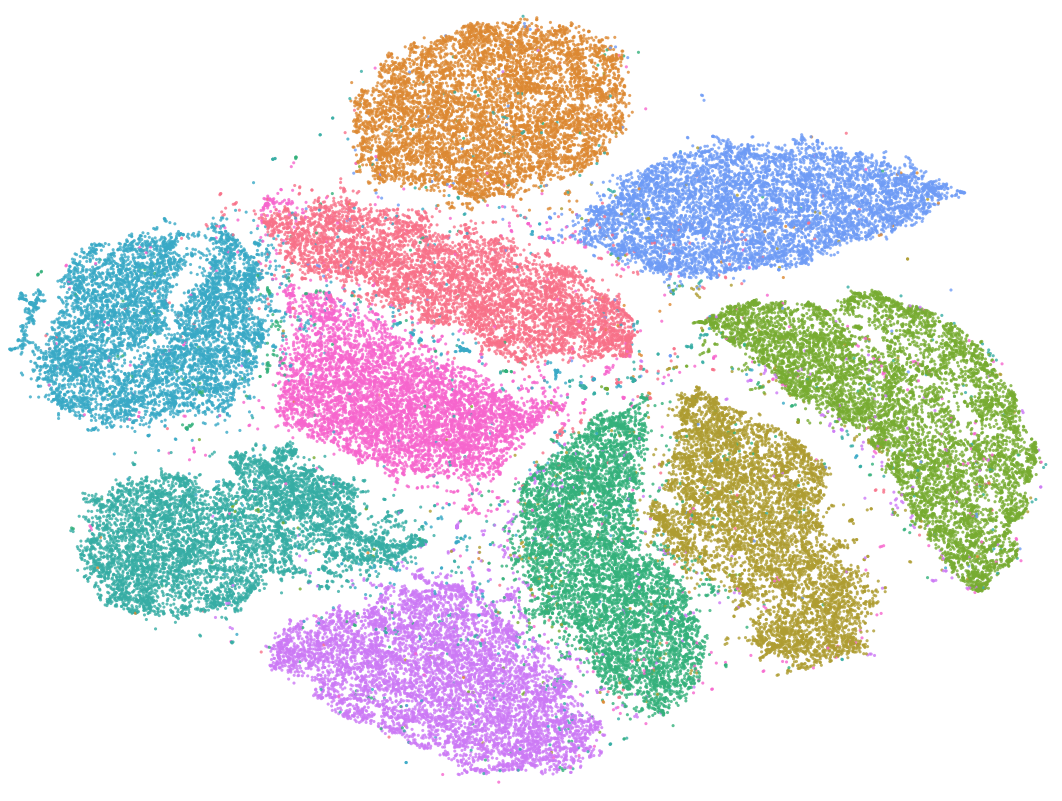} &   
   \includegraphics[width=0.3\linewidth]{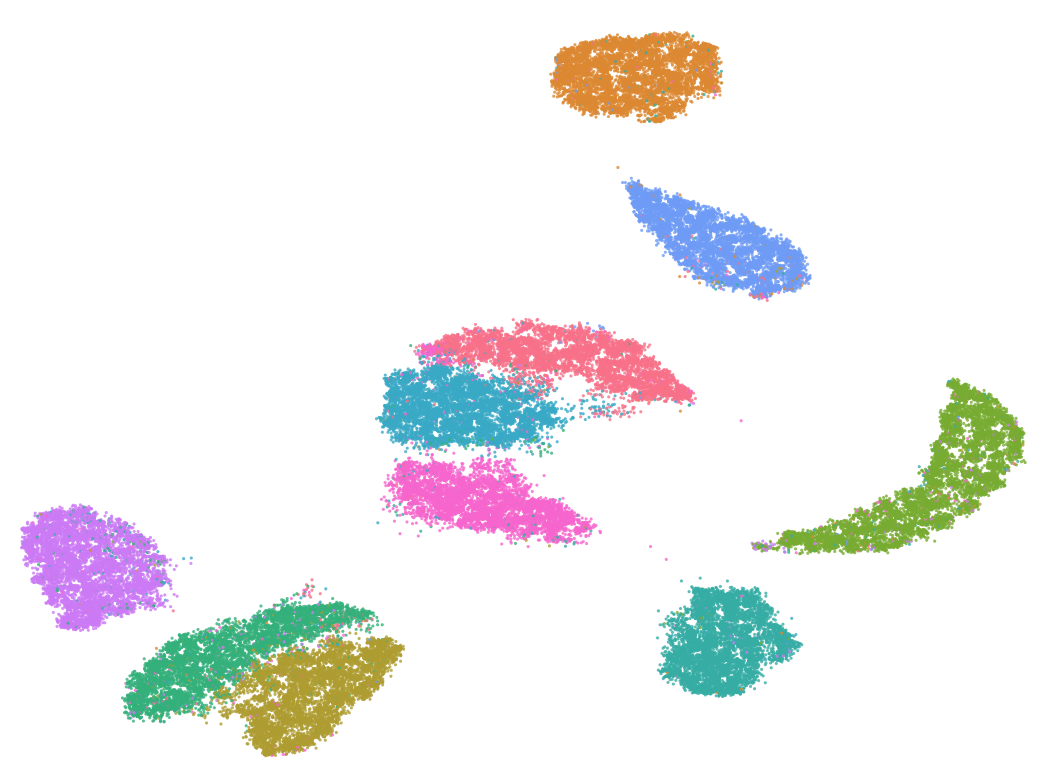} &
   \includegraphics[width=0.3\linewidth]{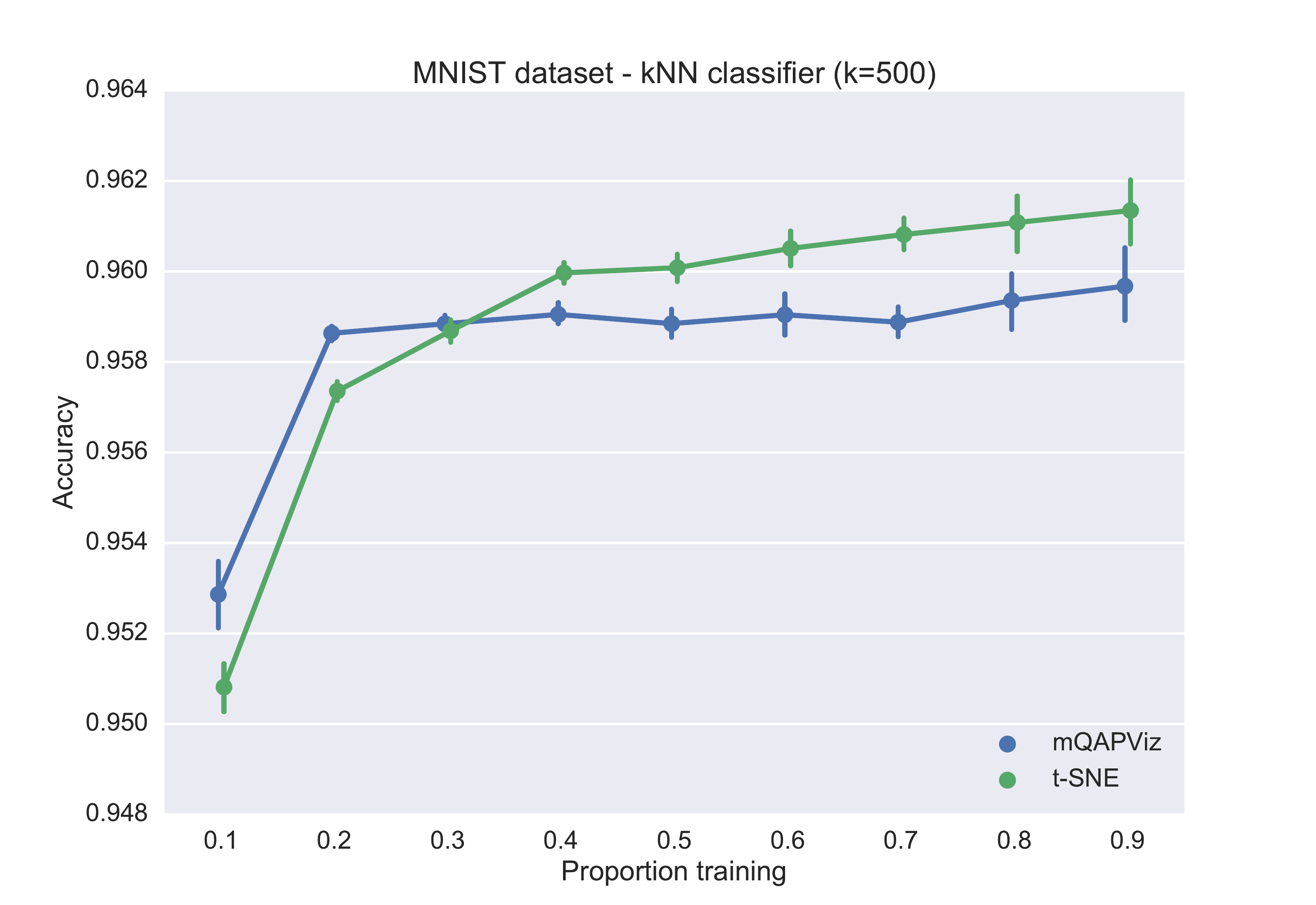} \\ 
   (d) MNIST (\acrshort{tSNE}) & (e) MNIST (\myapproach) & (f) MNIST (\acrshort{tSNE} vs. \myapproach)  \\[6pt] 
   \includegraphics[width=0.3\linewidth]{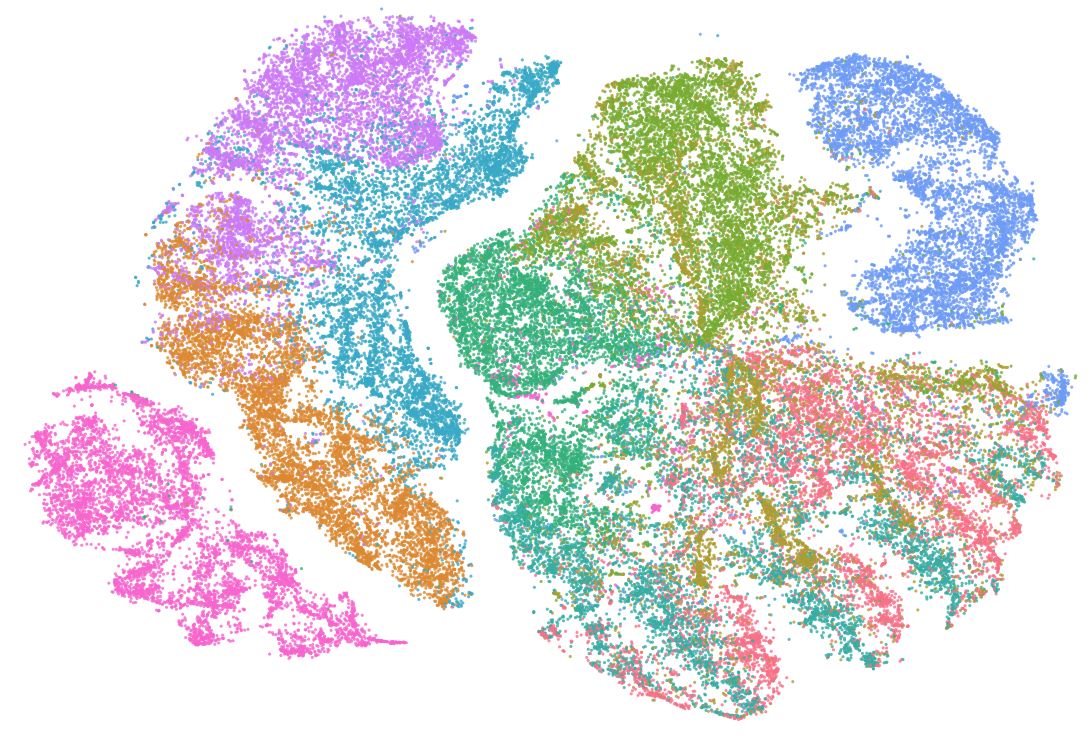} &   \includegraphics[width=0.3\linewidth]{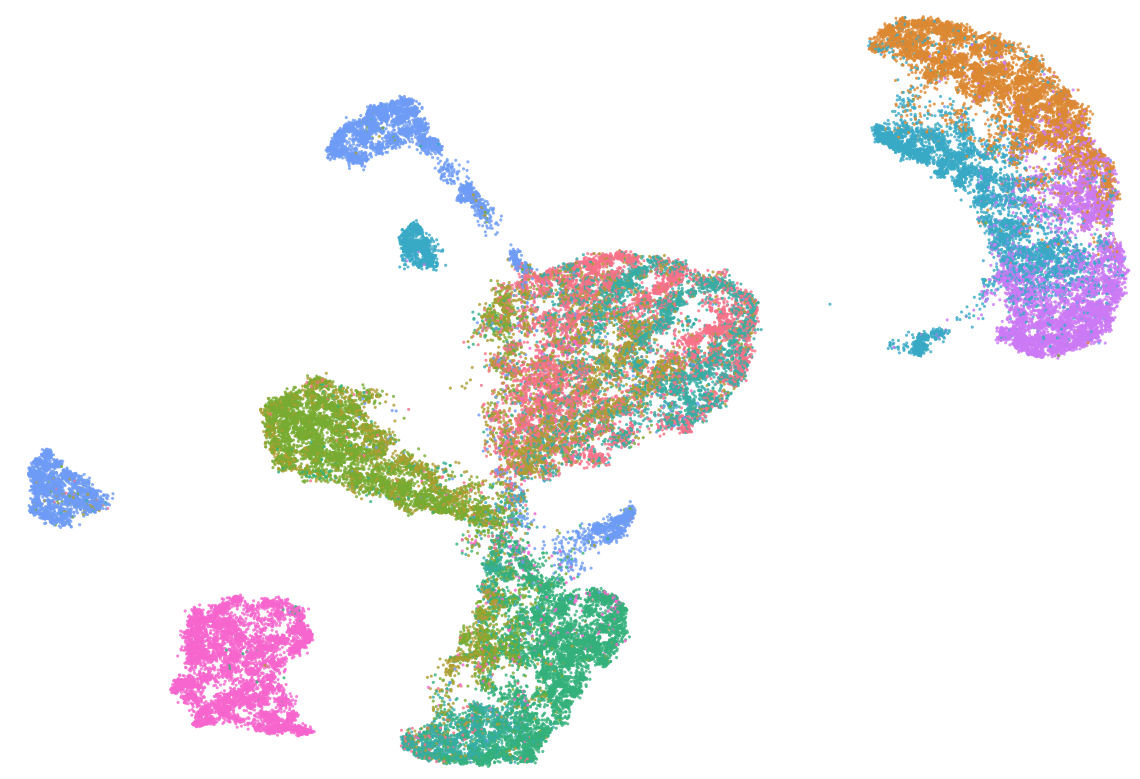} &
   \includegraphics[width=0.3\linewidth]{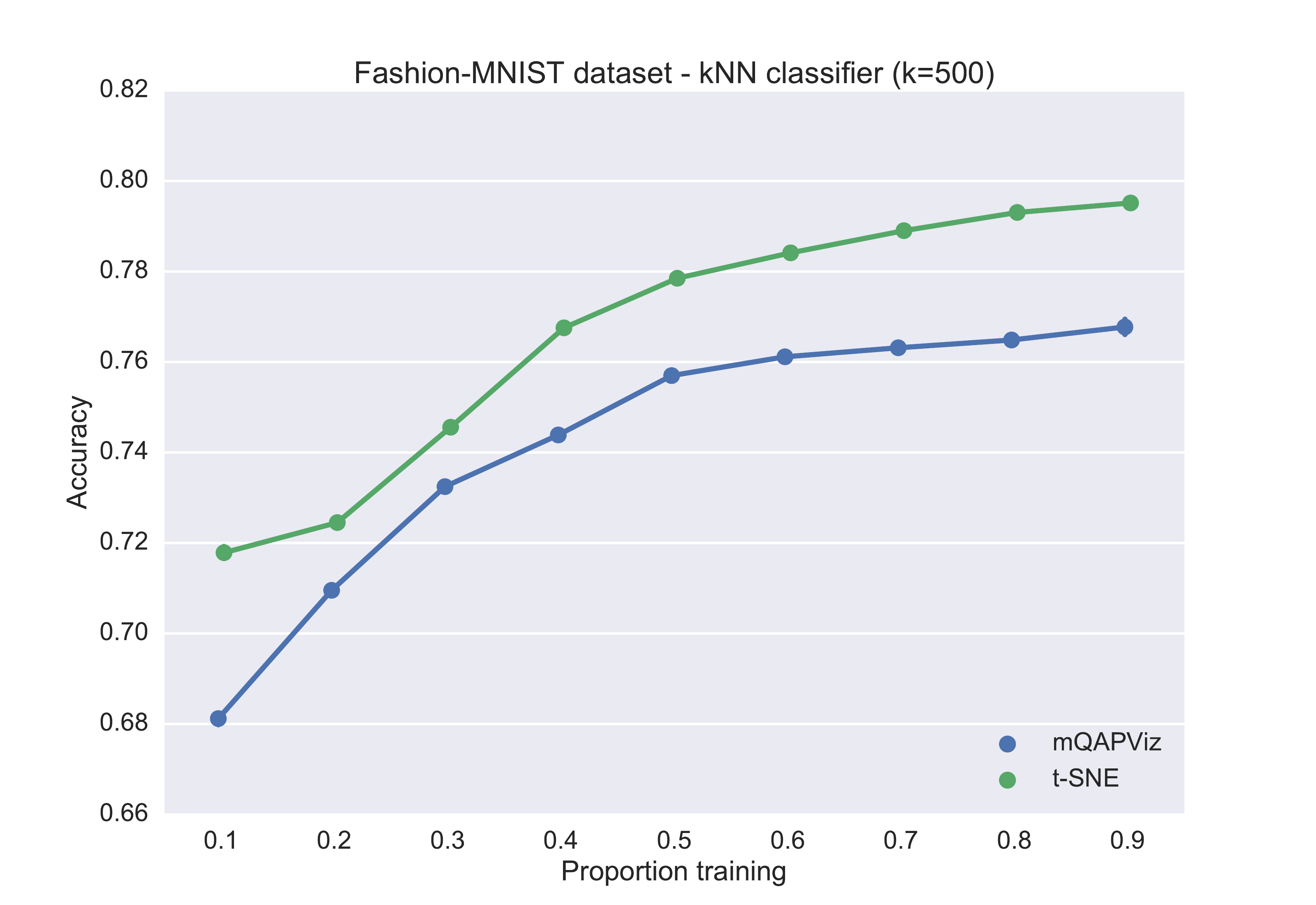} \\
   (g) Fashion-MNIST (\acrshort{tSNE}) & (h) Fashion-MNIST (\myapproach) & (i) Fashion-MNIST (\acrshort{tSNE} vs. \myapproach) \\[6pt]
   \includegraphics[width=0.3\linewidth]{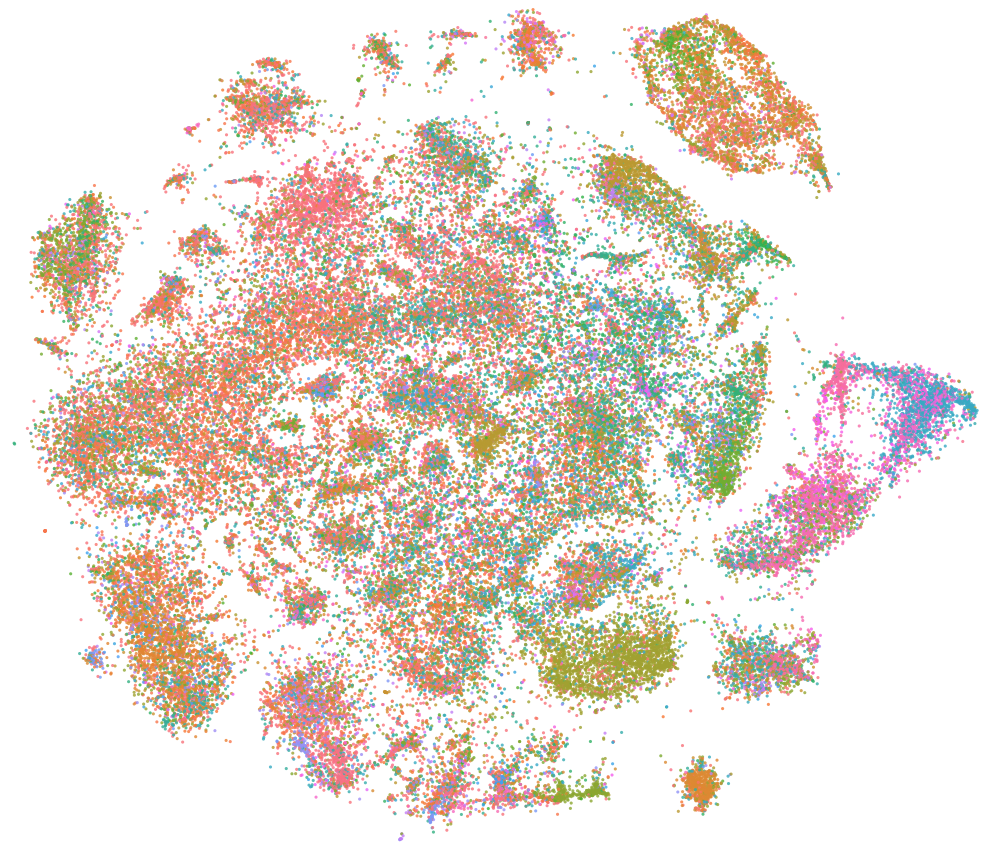} &   \includegraphics[width=0.3\linewidth]{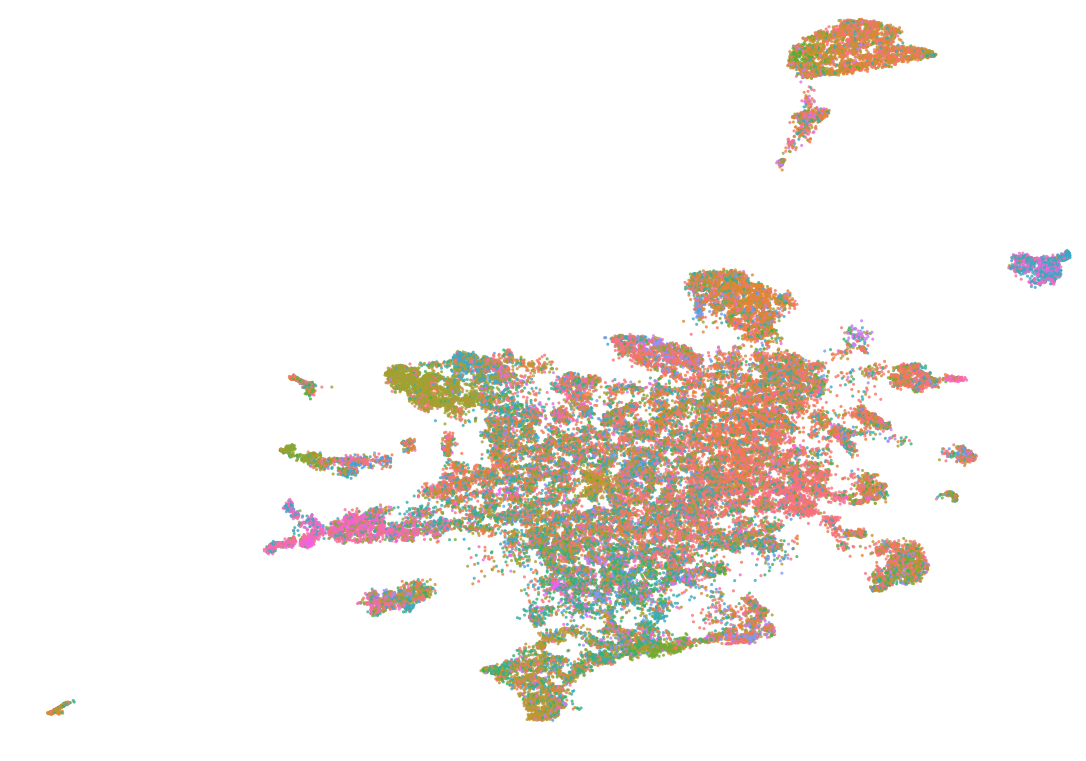} &
   \includegraphics[width=0.3\linewidth]{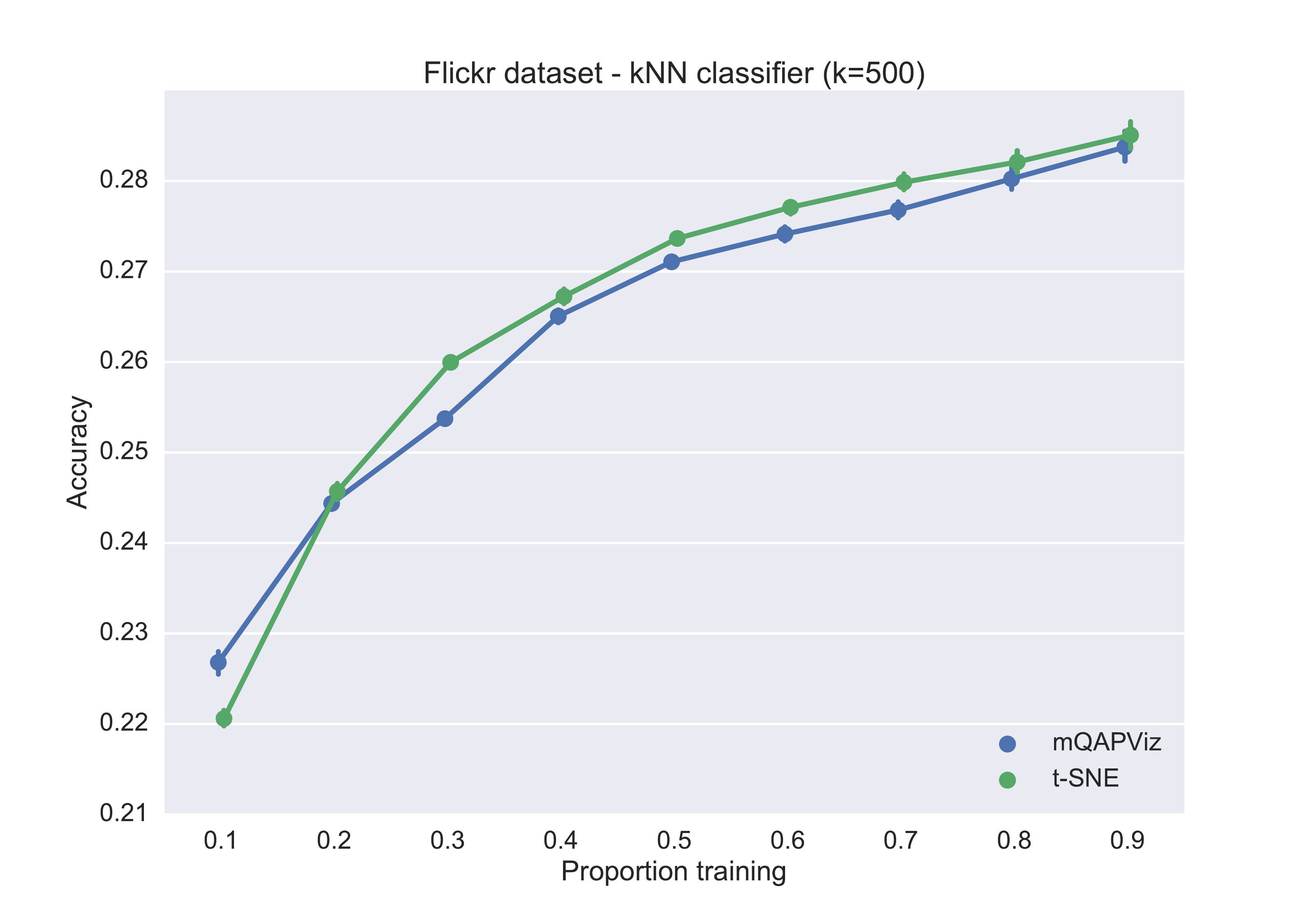}\\
   (j) Flickr (\acrshort{tSNE}) & (k) Flickr (\myapproach)  & (l) Flickr (\acrshort{tSNE} vs. \myapproach) \\[6pt]
 \end{tabular}
 \caption{Visualization of the datasets that contain true classes. Each color corresponds to a target class. We depict the visualizations obtained by t-SNE (column 1) and the visualizations obtained by \myapproach~(column 2). We also present the mean testing accuracy on different portions of training data. We executed 30 experiments on each training set, and we included the corresponding 95\% confidence interval (column 3).}
 \label{fig:visualizations}
 \end{figure*}

\begin{figure*}[ht]
	\centering
	\begin{tabular}{cccc}
		\includegraphics[width=0.23\linewidth]{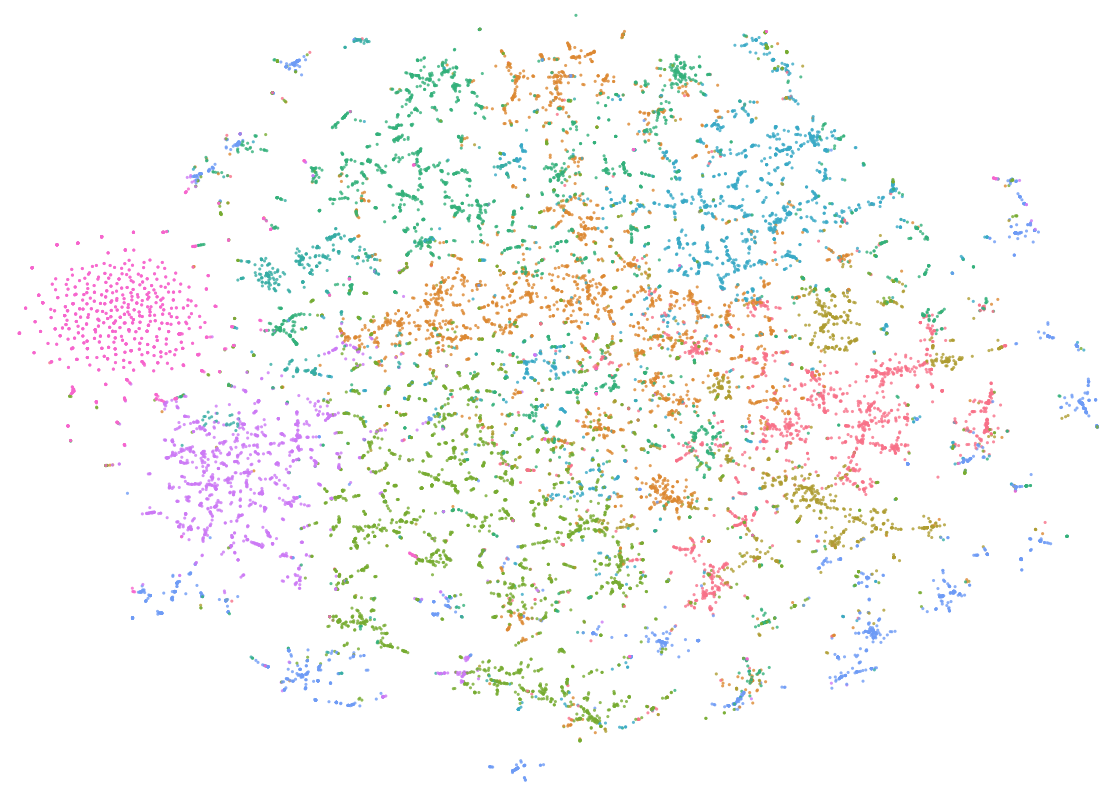} & 
		\includegraphics[width=0.23\linewidth]{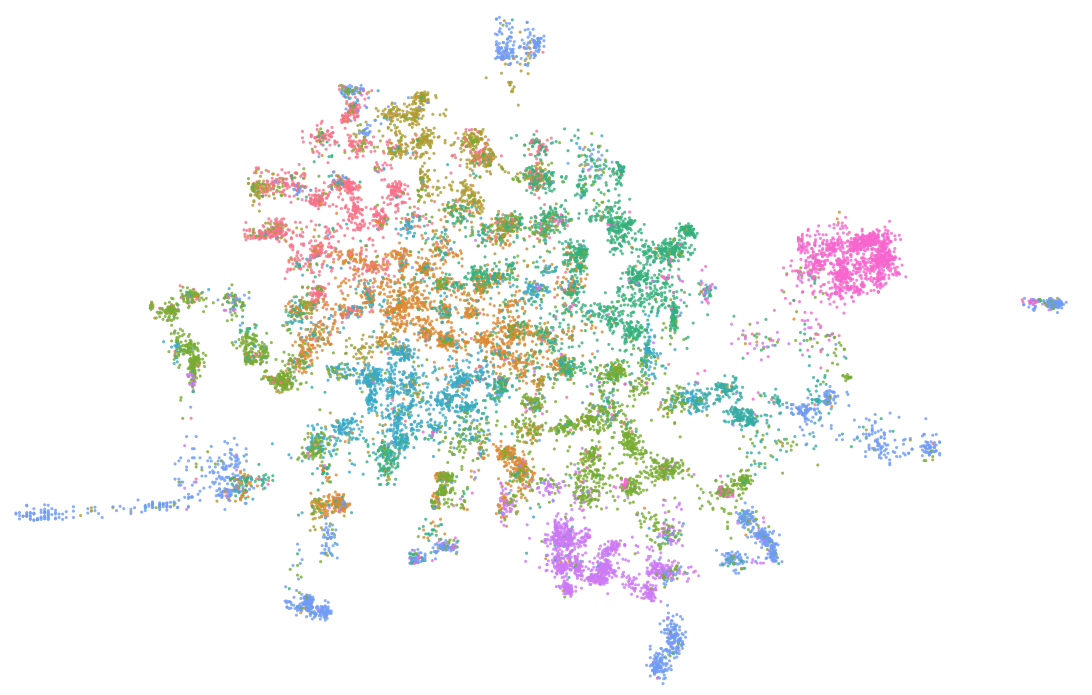} &
		\includegraphics[width=0.23\textwidth]{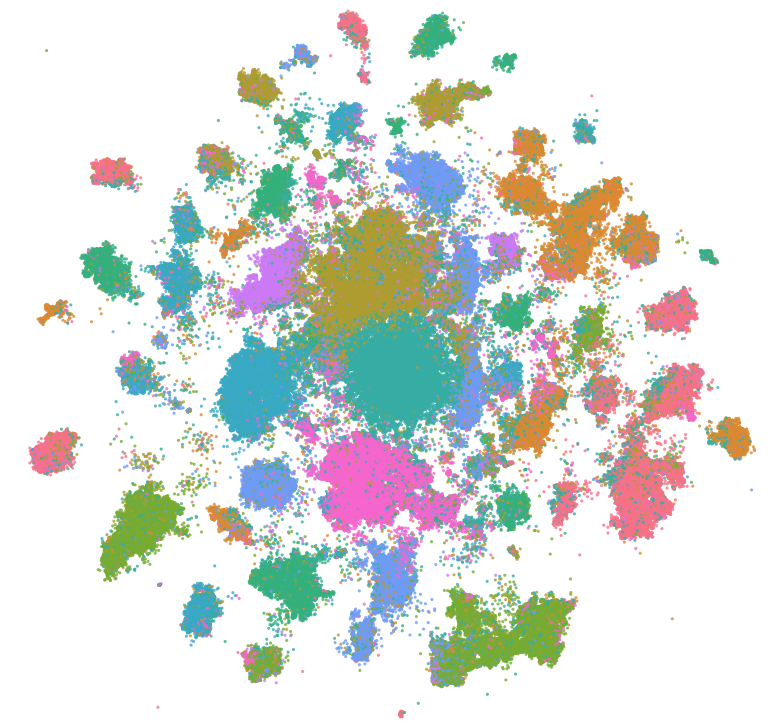} &
		\includegraphics[width=0.23\textwidth]{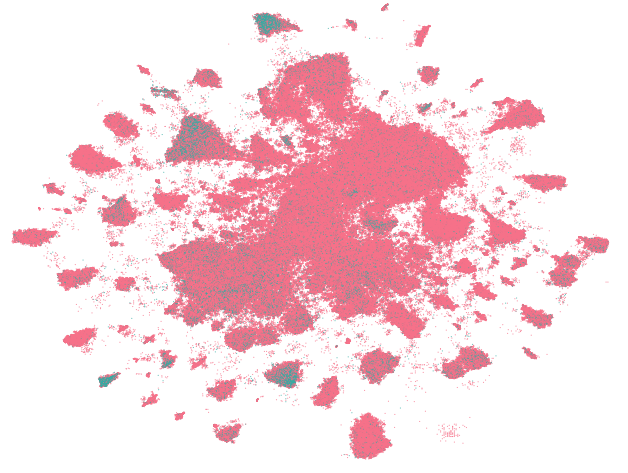} \\
	  (a) Astroph (\acrshort{tSNE}) & (b) Astroph (\myapproach) & (c) Pokec (\myapproach) & (d) Spammers (\myapproach) \\[6pt]
	\end{tabular}
\caption{Visualization computed using t-SNE and \myapproach. In the Astroph visualizations depicted in (a) and (b), and the Pokec visualization (c) we use the K-means algorithm to obtain ten groups represented by the colors. In the case of the Spammers dataset (d), we use the original labels \emph{spammer} (blue) and \emph{not spammer} (red).}
\label{fig:visualizations_large}
\end{figure*}

\subsection{Datasets} 
We evaluate \myapproach~with multiple real-world and large-scale datasets (\autoref{tab:experiments:datasets}).
In particular, we assess our method with the following datasets:
\begin{itemize}
\item \textbf{Astroph}: the Astro Physics collaboration network of authors who submitted papers to the Astro Physics category in arXiv\footnote{\url{http://snap.stanford.edu/data/ca-AstroPh.html}}. Each author is a data sample and an undirected edge corresponds to two authors that co-authored a publication.
\item \textbf{Pubmed}: the diabetes scientific publication network\footnote{https://github.com/jcatw/scnn/tree/master/scnn/data/Pubmed-Diabetes}. Each publication is a data sample and a directed edge represents that a publication cites another one.
\item \textbf{MNIST}: the handwritten digits dataset\footnote{\url{http://yann.lecun.com/exdb/mnist/}} in which each image is treated as a data object. 
\item \textbf{Fashion-MNIST}: the grayscale clothes dataset\footnote{\url{https://github.com/zalandoresearch/fashion-mnist}} in which each image is treated as a data object.
\item \textbf{Flickr}: the friendship network on Flickr\footnote{\url{http://socialcomputing.asu.edu/datasets/Flickr}}. Each user is a data object and an undirected edge represent the friendship between two users.
\item \textbf{Pokec}: the Slovakian social network dataset\footnote{\url{http://snap.stanford.edu/data/soc-pokec.html}}. Each user is a data object and an undirected edge represent the friendship between two users.
\item \textbf{Spammers}: the anonymized spammers social network dataset\footnote{\url{https://linqs-data.soe.ucsc.edu/public/social_spammer/}}. Each user is a data object which was manually labeled as \emph{spammer} or \emph{not spammer}. Given a user $u_i$ who performs an action targeting user $u_j$, a directed edge is created from $u_i$ to $u_j$.
\end{itemize}

Note that in the case of the network datasets, we first learn 
a feature vector representation for each node.
Although \textsc{DeepWalk} \cite{perozzi2014deepwalk} and \textsc{node2vec} \cite{grover2016node2vec} are two extremely efficient random walk-based representation learning algorithms, in our experiments we found that \textsc{LINE} \cite{tang2015line} performs better on the particular visualization task.
In consequence, we learn node representations through the \textsc{LINE} algorithm, and we represent each node by a vector of 128 dimensions, a value already proposed in the representation learning literature~\cite{tang2015line}.

\subsection{Evaluation}
To evaluate \myapproach, we compare our visualizations against the accelerated state-of-the-art approach for visualizing high dimensional data called \acrshort{tSNE}.
We use the C++ Barnes-Hut \acrshort{tSNE} implementation published by the authors\footnote{\url{https://lvdmaaten.github.io/tsne/}}.

\textbf{Model parameters and settings -- }
For the model parameters in \acrshort{tSNE}, we set $\theta = 0.5$, the number of iterations to 1,000, and the initial learning rate to 200 which are suggested in \cite{maaten2014accelerating}.
For both \textsc{LINE} and \textsc{LargeVis}, the size of mini-batches is set as 1; the learning rate is set as $\rho_t = \rho(1-t/T)$, where $T$ is the total number of edges samples or mini-batches.
The initial learning rates used by \textsc{LINE} and \textsc{LargeVis} are $\rho_0 = 0.025$ and $\rho_0 = 1$ respectively. 
All these parameters are suggested by the authors of \cite{tang2016visualizing} (including setting the number of negative samples to 5 and the unified weight of the negative edges to 7).
In \myapproach, we compute the visualizations by sampling 30\% of the nodes in the $G_{k-nn}$.

\textbf{Quantitative evaluation -- }
Assessing the quality of a visualization outcome is an inherently subjective task.
To overcome this issue and to quantitatively evaluate the visualizations, we apply the $k$-NN classifier (implemented in scikit-learn\footnote{\url{http://scikit-learn.org}}) to classify the samples based on their visualization outcomes (i.e., 2D representation).
The idea of this methodology is that a good visualization should be able to preserve the structure of the original data as much as possible and, therefore, a high classification accuracy would still be present even if just working with the low-dimensional representation.
We report on the results of $k$-NN classifiers on different proportions of the training data (with $k = 500$).
For each proportion, we train the classifier thirty times on different training sets.
To evaluate the performance of a classifier, we report the mean testing accuracy over the thirty rounds and the corresponding 95\% confidence interval (column 3, \autoref{fig:visualizations}).
We observe that in three out of four datasets, \myapproach~is at least competitive with respect to \acrshort{tSNE}.
Fashion-MNIST is the only dataset in which we can see that \acrshort{tSNE} quantitatively outperforms \myapproach.

\textbf{Visualizations -- }
We show several visualization examples to evaluate the quality of \myapproach~visualizations against \acrshort{tSNE} (\autoref{fig:visualizations}, columns 1 and 2).
The colors correspond to the classes (Pubmed, MNIST, Fashion-MNIST, Flickr, and Spammers) or partitions computed with the K-means based on the high-dimensional representation (Astroph and Pokec, \autoref{fig:visualizations_large}) in which we partitioned the dataset in ten groups.
We observe in the smallest dataset that the visualizations generated by both methods are meaningful and comparable to each other.
On the larger datasets, we argue the visualizations generated by the \myapproach~are more intuitive.
In the case of our larger datasets with 1.6M and 5.3M objects, \acrshort{tSNE} could not compute a visualization due to its high memory consumption.
We can see in the Pokec visualization computed with \myapproach~(\autoref{fig:visualizations_large}c) several groups of objects that seem to share some common characteristics.
In the case of the Spammer visualization (\autoref{fig:visualizations_large}d), we observe that spammer users are grouped in, at least, four different regions of the layout.
With ad-hoc tools, we may isolate these lands of objects to perform further analyses.

\section{Conclusions}
\label{sec:conclusions}
In this study we proposed what is, to the best of our knowledge, the first method that uses a multi-objective optimization algorithm to compute visualizations of large datasets.
\myapproach~is based on a divide-and-conquer approach in which several \acrshort{mQAP} sub-instances are defined using the layout induced by sampled nodes and their nearest neighbors that belong to an efficiently computed \acrshort{KNNG}.
Although we report results on a cluster grid, the method also allows us to generate visualizations of a million data objects in a single multi-core machine without requiring any special distributed processing architecture.
Our experiments showed that \myapproach~is competitive against \acrshort{tSNE} using a simple quantitative evaluation.
The visualizations generated with \myapproach~can later be used for further data analyses tasks. 

We limited our study by evaluating two objective functions.
However, the method could also accept other alternative objective functions, for example, considering both high- and low-dimensional data.
Also, at this moment, our method can be used on \emph{data snapshots}, excluding potential temporal associations between the objects.
Thus, another challenging direction is to compute visualizations using multi-objective optimization to support the analysis of datasets that dynamically change across space and time (i.e., spatio-temporal datasets).

\begin{acks}
C.S. and F.J. are funded by the UNRSC50:50 Ph.D. scholarship at The University of Newcastle.
P.M. and R.B. acknowledge previous funding of their research by the Australian Research Council (ARC) Discovery Project DP140104183.
P.M. also acknowledges ARC support with his Future Fellowship FT120100060.
The authors are grateful to Aaron Scott for his IT support with the University's HPC architecture.
\end{acks}

\bibliographystyle{ACM-Reference-Format}
\bibliography{00_mqapviz} 

\end{document}